\documentclass[aps,prd,twocolumn,showpacs,preprintnumbers,a4paper]{revtex4}
\usepackage{dcolumn}
\usepackage{bm}
\usepackage{amssymb,amsmath}
\usepackage[english]{babel}
\usepackage[dvips]{graphicx}
\usepackage{subfigure}
\usepackage{color}
\usepackage{psfrag}

\newcommand{\df}{\delta\varphi}
\newcommand{\ddf}{\:\:\dot{\!\!\delta\varphi}}

\newcommand{\dfk}{\delta\varphi_\mathbf{k}}

\newcommand{\dchik}{\delta\chi_\mathbf{k}}
\newcommand{\xik}{\xi_\mathbf{k}}

\renewcommand{\t}{\tau}
\newcommand{\tilt}{\tilde\tau}

\newcommand{\tin}{\tau_{in}}
\newcommand{\s}{\sigma}

\newcommand{\spect}{\mathcal{P}_{\df}(k)}

\newcommand{\mean}[1]{\left\langle #1 \right\rangle}

\newcommand{\dx}{d^4\!x}
\newcommand{\dk}{d\mathbf{k}}
\newcommand{\de}{\partial}

\newcommand{\eps}{\epsilon}
\newcommand{\al}{\alpha}

\newcommand{\Ainv}{\mathbf{A}^{-1}}

\newcommand{\re}{\mathrm{Re}}

\newcommand{\kk}{\mathbf{k}}
\newcommand{\xx}{\mathbf{x}}

\newcommand{\rr}{ {\bf r}}

\begin{document}
\thispagestyle{empty}

\title{Stochastic Inflation and the Lower Multipoles in the CMB Anisotropies}

\author{Michele Liguori}
\email{michele.liguori@pd.infn.it}
\author{Sabino Matarrese}
\email{matarrese@pd.infn.it}
\affiliation{Dipartimento di Fisica `G. Galilei', Universit\`{a} di Padova,
\& INFN - Sezione di Padova, via F. Marzolo 8, I-35131 Padova, Italy}
\author{Marcello A. Musso}
\email{marcello.musso@pv.infn.it}
\affiliation{Dipartimento di Fisica Nucleare e Teorica,
Universit\`{a} degli Studi di Pavia \\\& INFN - 
Sezione di Pavia, via U. Bassi 6, I-27100, Pavia, Italy}
\author{Antonio Riotto}
\email{antonio.riotto@pd.infn.it}
\affiliation{INFN - Sezione di Padova, via F. 
Marzolo 8, I-35131 Padova, Italy}

\date{\today}

\begin{abstract}
\noindent
We generalize the treatment of inflationary perturbations to 
deal with the non-Markovian colored noise emerging from any realistic 
approach to stochastic inflation. 
We provide a calculation of the power-spectrum of the 
gauge-invariant comoving curvature perturbation to first order 
in the slow-roll parameters. Properly accounting for the constraint that 
our local patch of the Universe is homogeneous on scales just above the 
present Hubble radius, we find a blue tilt of the power-spectrum
on the largest observable scales, in agreement with the \emph{WMAP} data 
which show an unexpected suppression of the low multipoles of the CMB 
anisotropy. Our explanation of the anomalous behaviour of the 
lower multipoles of the CMB anisotropies does not invoke  
any {\it ad-hoc} introduction of new physical ingredients in 
the theory.

\end{abstract}


\preprint{DFPD/04/A-13}

\maketitle

\numberwithin{equation}{section}


\section{Introduction}

During the last two decades the inflationary paradigm has become an almost 
universally accepted scenario to explain the observed large scale flatness 
and homogeneity of the Universe \cite{lrreview, tonireview}. 
At the same time, it provides an
efficient mechanism (the stretching of quantum fluctuations up to cosmological
scales) to generate those small curvature perturbations (of the order 
of $10^{-5}$) that are believed to produce the 
observed Cosmic Microwave Background (CMB) temperature anisotropies and to 
provide the seeds for the formation of the cosmic structures 
(galaxies and clusters) we observe today. 

Stochastic inflation \cite{early} represents one possible approach to the 
inflationary paradigm, specifically conceived to describe the transition 
from quantum fluctuations to classical density perturbations 
\cite{decoherence}. The basic idea is to introduce a cutoff
in Fourier space through a suitable time-dependent window function that
filters out the modes whose frequency is lower than the comoving horizon 
size. 
The inflaton field is thus split in two differently behaving parts:
the short-wavelength part has an intrinsically quantum nature, while the 
coarse-grained one (collecting the remaining super-horizon modes) is 
treated as classical. A 
Langevin-like equation of motion for the long-wavelength 
part is obtained, where the sub-horizon modes enter now as a classical 
stochastic noise term that perturbs the dynamics of the coarse-grained
field.

In most works on this subject \footnote{The stochastic approach
to inflation has been used in Refs. \cite{constraint,stochNG} to
address the issue of non-linear inflationary perturbations.} the noise is 
taken to be a 
\emph{white noise}, whose two-point correlation function is by definition 
proportional to a Dirac delta function in time. 
A remarkable feature of stochastic processes
involving a white noise, known
as Markov property, is that their conditional probability does not depend
on times before the constraint. For this particular kind of noise,
standard techniques of statistical physics \cite{fokkerplanck} allow to 
write an evolution equation (the Fokker-Planck equation) for the probability
distribution of the inflaton perturbations. 

A key issue in the solution of the Langevin or Fokker-Planck equation is 
the choice of the initial conditions for the perturbations. Many authors 
\cite{constraint} agree that it should be consistent to 
assume the spatial homogeneity of our observable local patch of the Universe,
and therefore the vanishing of all fluctuations right before the moment it
crosses the horizon size, about 60 e-folds before the end of inflation,
since at that time only fluctuations on larger scales could have grown up.
Therefore, all points inside the present Hubble radius (at that time
contained in the same coarse-graining domain) must have the same local 
value of the scalar field, although this value can be different from
the one assumed in other regions of the Universe.
Even if it is generally assumed that inflation started 
well before the last 60 e-folds, for the white-noise case the evolution 
of fluctuations is completely insensitive to what happened before that 
epoch and the constraint really becomes a new initial condition.

However, a white noise arises only if the window function is chosen to
be a step-function in Fourier space, therefore if coarse-graining is achieved 
with the introduction of a sharp momentum cutoff.
Even though it must be regarded as the mathematical limit of 
a smoother filtering procedure, it has been pointed out \cite{colored1}
that the step function is a somewhat pathological choice, because 
it does not satisfy some fundamental properties characterizing
``well-behaved'' smooth window functions. Conversely, all these window 
functions always produce a non-divergent colored noise with the same 
asymptotical behaviour.
Unfortunately, the statistical treatment of a colored noise 
is a rather more difficult matter \cite{colored2}, but in general it
is expected that because of their non-Markovianity fluctuations 
will keep some memory of the evolution before the constraint.

In a previous paper \cite{MMR}, we studied the dynamics of the fluctuations
of the coarse-grained inflaton field in a pure de Sitter space-time
and found evidence for a blue tilt in the power-spectrum on the 
largest observable scales as a consequence of the non-Markovian dynamics 
near the constraint. This is due to the fact that the increased noise
correlation time (with respect to the white-noise case) acts as a sort of
``inertia'' against the growth of the perturbations after the constraint,
thereby resulting in a suppression of the power-spectrum on the scales that 
crossed the horizon in the following few Hubble times.

This is an interesting feature, since the CMB anisotropy measurements 
made by \emph{WMAP} \cite{WMAP} give some evidence for a suppression of the 
low multipoles, confirming earlier analogous results found by \emph{COBE}
\cite{COBE}. Although the statistical significance of such a suppression is 
not large \cite{lowmultstat}, many authors have recently tried to
explain this spectral feature either invoking astrophysical effects 
\cite{blue-tilt1} or introducing some new physical input  in the
mechanism that generates the perturbations \cite{blue-tilt2}.

In this paper, we address the question of the low multipoles suppression
from the stochastic inflation point 
of view, suggesting that this might be simply a consequence of
the colored noise. Compared to a white noise, a smooth 
choice of the window will in fact slightly suppress the contribution to the
noise given by the field modes whose frequency is immediately higher than 
the cutoff scale $\sigma (a H)$ (while enhancing lower frequencies) 
where $\sigma$ is a parameter smaller than unity, which
is introduced to parametrize to  level of arbitrariness 
-- besides the shape of the window function -- 
in the size of the coarse-graining 
domain. Right after the time $\t_*$ at which we set the homogeneity 
constraint on our comoving patch of Universe, 
fluctuations with $k\lesssim\sigma a_*H_*$ 
will grow less than in the white-noise case before freezing out,
and if $\sigma$ is not too small this suppression can be effective
also on observable scales.
However, there seems to be no physical motivations to state that the 
coarse-graining domain has to be very large compared to the horizon size.
This assumption indeed appears in early stochastic inflation works, 
where $\sigma$ is taken to be much 
smaller than unity, but this is done essentially for practical reasons, in 
order to ease up the calculations and to cast away from observable scales 
the possible effects 
introduced by the white-noise approximation. This choice is justified 
\emph{a posteriori} with the argument that the noise correlation function 
in configuration space is independent of $\sigma$ up to second order. 
Nevertheless, a $\sigma$ dependence still remains (even in the Markovian
case) in the power-spectrum.

The plan of the paper is as follows: in Section \ref{effaction}
we briefly review the formalism of stochastic inflation, extending it
to the colored-noise case \cite{MMR}. In Section \ref{fluctuations} we exactly 
calculate the colored-noise correlation function for a free 
massive scalar field in a de Sitter space-time.
In Section \ref{probabdistr} we evaluate the effects of
the constraint at $\t_*$ computing the conditional probability distribution
of the same scalar field. In Section \ref{curvature} we extend our
results to the perturbations of the inflaton field (including in our 
treatment also metric fluctuations) and we compute the power-spectrum 
$\mathcal{P}_\mathcal{R}$ of the gauge-invariant 
curvature perturbation, taking as the only approximation a 
first-order slow-roll expansion of the background inflaton field.
Finally, our conclusions are drawn in Section \ref{conclu}.


\section{Effective stochastic action}
\label{effaction}

We begin our considerations by taking a free scalar field with 
mass $m$ in a homogeneous and isotropic background space-time, 
whose metric has the pure Robertson-Walker form
\begin{equation}
\label{desitter}
  ds^2=dt^2-a^2(t)d\mathbf{x}^2=a^2(\t)(d\t^2-d\mathbf{x}^2),
\end{equation} 
where $a(t)$ is the scale factor responsible for the expansion of the Universe
and $\t$ is the conformal time, defined in such a way that $d\t=dt/a(t)$.

The ordinary action for a scalar field in a curved space-time 
with this background metric reads 
\begin{equation}
\label{azione}
  S[\phi]=\int\!\!\dx 
a^3(t)\frac{1}{2}\!\left[(\de_t\phi)^2-\frac{(\nabla\phi)^2}
{a^2}-m^2\phi^2\right],
\end{equation}
and the equation of motion for the normal modes $\phi_\kk$ (assuming 
a spatially flat Universe these are simply obtained by expanding the 
field on the 3-dimensional Fourier basis $e^{i\kk\cdot\xx}/(2\pi)^{3/2}$) is
\begin{equation}
\label{eqclass}
  \ddot\phi_\kk+3H\dot\phi_\kk+\frac{k^2}{a^2}\phi_\kk+m^2\phi_\kk=0,
\end{equation}
where dots denote ordinary time derivatives.

It is convenient to define a new field variable $\chi$, 
whose normal modes are related to the Fourier modes $\phi_\kk$ by
\begin{equation}
  \chi_\kk=\phi_\kk a(\t);
\end{equation}
in conformal time, indicating with a prime the conformal-time derivative 
$\partial_\t=a\partial_t$, the equation of motion for the new
field becomes
\begin{equation}
\label{confeqclass}
  \chi_\kk'' + \left[k^2 + m^2a^2 - \frac{a''}{a}\right]\chi_\kk=0,
\end{equation}
and can be obtained by variation of the conformal action
\begin{equation}
  S[\chi]=-\int\!\dx \frac{1}{2}\,\chi\left[\square 
  + m^2a^2 - \frac{a''}{a}\right]\chi,
\end{equation}
where the metric determinant does not appear and 
$\square=\partial_\t^2-\nabla^2$ is the ordinary 
flat space-time d'Alembert operator.

In the de Sitter case, the Hubble parameter $H\equiv\dot a/a$ measuring the 
expansion rate is strictly constant in time, and the scale factor evolves as
\begin{equation}
  a(t)=e^{Ht},
\end{equation}
while the conformal time, ranging from $-\infty$ to 0, becomes
\begin{equation}
  \t=-\frac{1}{a(t)H}.
\end{equation}
The evolution equation for the $\chi_\kk$'s can then be recast in the form 
\begin{equation}
\label{dSeq}
  \chi_\kk'' + \left[k^2 - 
  \frac{1}{\t^2}\left(\nu^2-\frac{1}{4}\right)\right]\chi_\kk=0,
\end{equation}
where the parameter $\nu$ is defined as
\begin{equation}
  \nu=\sqrt{\frac{9}{4}-\frac{m^2}{H^2}} \equiv \frac{3}{2} - \eps\,.
\end{equation} 

The generic solution to this equation, expressed in terms of 
Bessel functions of the first and second kind, is 
\begin{equation}
  c_1\sqrt{|\t|}J_\nu(k|\t|) + c_2\sqrt{|\t|}Y_\nu(k|\t|);
\end{equation}
requiring each $\chi_\kk$ to match the plain wave solution 
$e^{-ik\t}/\sqrt{2k}$ for $k\gg aH$, when wavelengths are too short to 
feel any space-time curvature effects, produces the standard Bunch-Davies 
solution 
\begin{equation}
  \chi_\mathbf{k}(\t)=\frac{\sqrt{\pi}}{2} \sqrt{|\t|}H_\nu^{(1)}(k|\t|),
\end{equation}
where 
\begin{equation}
  H_\nu^{(1)}(x) = J_\nu(x)+iY_\nu(x)
\end{equation}
it the Hankel function of the first kind.
For the normal modes of the original scalar field $\phi$, this gives
\begin{equation}
  \phi_\mathbf{k}(\t)=\frac{\sqrt{\pi}}{2} H |\t|^{3/2}H_\nu^{(1)}(k|\t|),
\end{equation}
that in the massless case ($\nu=\frac{3}{2}$) becomes
\begin{equation}
  \label{mnormali}
  \phi_\mathbf{k}(\t)=H\frac{k\t-i}{\sqrt{2k^3}}
  e^{-ik\t}.
\end{equation}

Each mode is said to leave the horizon when the corresponding physical 
wavelength $ak^{-1}$ encompasses, due to the growth of the scale factor $a(t)$,
the size of the cosmological particle horizon $H^{-1}$. At the horizon 
crossing we therefore have $|k\t|=1$.

It is possible to obtain the sub-horizon part of the field selecting  
the short-wavelength modes, through a suitable time-dependent high-pass 
filter in Fourier space. We do this by means of a window function 
$W_\sigma(k\t)$ such that $W_\sigma(k\t)=0$ for $k|\t|\ll \sigma$ and 
$W_\sigma(k\t)=1$  for $k|\t|\gg \sigma$, where the parameter $\sigma$
(in early stochastic inflation works called $\eps$) introduces some level 
of arbitrariness in the size of the coarse-graining domain, corresponding
to the ``effective horizon'' $\sigma (aH)$. 
The fine-grained sub-horizon field $\phi_>$ is therefore defined as
\begin{equation}
  \phi_>=\!\int\!\! \dk \frac{W_\sigma(k\t)}{(2\pi)^{3/2}}
  \big[a_\kk\phi_\kk(\t)e^{-i\kk\cdot\xx} + h.c. \big],
\end{equation}
while the coarse-grained super-horizon part $\phi_<\equiv\varphi$ is left 
for the moment unspecified.

An effective equation of motion for the coarse-grained part can simply 
be obtained by substituting $\phi=\varphi+\phi_>$ directly in the equation 
of motion \eqref{eqclass} \cite{early}; all the high-frequency part is
then collected into a random noise field which acts as a classical source
term for the long-wavelengths part. The quantum fluctuations on sub-horizon 
scales are averaged into a classical noise term $\xi$ (with a given
probability distribution $P[\xi]$) perturbing the super-horizon 
dynamics: the effects of such a perturbation  is then a stochastic process
to be studied with the ordinary methods of classical statistical physics.
The quantum problem of computing the expectation value of the coarse-grained
field $\varphi$ is thus reduced to the classical problem of evaluating the
mean of the solution to the stochastic evolution equation averaged over
all possible noise configurations.

However, things are a bit more complicated than this, 
since we are dealing with true expectation
values between the same \emph{in} state, and not simply with \emph{in-out} 
transition amplitudes (which are the usual objects in ordinary quantum 
field theory). The calculation of expectation values is a typical problem of 
out-of-equilibrium field theory \cite{noneqFT}.
In this framework, an effective action for $\mean{\varphi}$ can be obtained
via the so-called influence functional method \cite{influence}, where
the splitting of the field is operated in the action, and the short 
wavelengths are integrated out with a path-integral over all the 
configurations of the sub-horizon field $\phi_>$. This method
actually introduces some extra terms in the effective equation of 
motion \cite{MMR}.
However, these effects are small and are neglected in the present work.

Since we take the effect of sub-horizon fluctuations on large 
scales to be small, we can split the field into its statistical mean value 
$\varphi$ (satisfying the classical equation of motion \eqref{eqclass}) 
and a fluctuation $\df[\xi]$, that by 
definition vanishes when averaged over the $\xi$'s. 
The stochastic equation of motion for the super-horizon fluctuations 
can be reduced to
\begin{gather}
\label{eqmoto}
  \delta\ddot\varphi_\kk+3H\ddf_\kk-\bigg(\frac{k^2}{a^2}-m^2\bigg)
  \dfk = \frac{\xi_\kk}{a^3},
\intertext{or equivalently}
\label{eqmotoconf}
  \dchik''+\bigg(k^2 -\frac{a''}{a} + m^2a^2\bigg)\dchik = \xik,
\end{gather}
which are the usual equations obtained in stochastic inflation inserting the
frequency split of the field directly into the equation of motion. This is
a fully classical Langevin-like equation, where the noise $\xi$ 
(resulting from the average of the quantum fluctuations on sub-horizon 
scales) stochastically perturbs the super-horizon dynamics, acting as a 
source for the long-wavelengths field $\varphi$, which in turn can be treated
as a classical stochastic variable.

The noise $\xi$ is a Gaussian random field, whose configurations are 
weighted by the functional probability distribution
\begin{align}
\label{weight}
  P[\xi] &= N\exp\!\left[
  -\frac{1}{2}\!\int\!\!\dx\dx'\xi(x)\Ainv(x,x')\xi(x')\right]\\
  &= N\exp\!\left[-\frac{1}{2}\!\int\!\!d\t d\t'\dk\dk'
  \xik(\t)\mathbf{A}_{\kk,\kk'}^{-1}(\t,\t')\xik(\t')\right], \nonumber
\end{align}
where $\mathbf{A}_{\kk,\kk'}^{-1}(\t,\t')$ is the functional inverse of
\begin{align}
\label{noisecorr}
  \mathbf{A}_{\kk,\kk'}(\t,\t') 
  =\delta(\kk+\kk') \frac{\re [f(k\t)f^*(k\t')]}{2k^3} ,
\end{align}
and
\begin{align}
  f(k\t)= \sqrt{2k^3}(W_\sigma''\chi_\kk+2W_\sigma'\chi'_\kk).
\end{align}

This probability distribution allows us to 
calculate the statistical mean value $\mean{\dots}_S$ of any 
$\xi$-dependent quantity averaged over
all the noise field configurations, defined as
\begin{equation}
\label{mean}
  \mean{\dots}_S=\int\!\!\mathcal{D}[\xi]\dots P[\xi].
\end{equation}
Then, by definition the mean $\mean{\xi(\t)}_S$ of the noise vanishes at
all times, while the two-point correlation function, which completely 
characterizes the statistical properties of the Gaussian noise field, 
is then by definition
\begin{equation}
  \mean{\xi_\kk(\t)\xi_{\kk'}(\t')}_S =\mathbf{A}_{\kk,\kk'}(t,t') 
\end{equation}
In configuration space the correlation function reads
\begin{multline}
  \mean{\xi(x)\xi(x')}_S = \!\int\!\!\frac{\dk}{(2\pi)^3} 
  e^{i\kk(\xx-\xx')} \frac{1}{2k^3} \\ \re [f(k\t)f^*(k\t')].
\end{multline}

The statistical behaviour of the noise thus critically depends on the 
shape of the filter. Choosing the usual window  
$W_\sigma(k\t)=\vartheta(k|\t|-\sigma)$ we obtain the standard white-noise 
two-point correlation function. For $\xx=\xx'$ it reads
\begin{equation}
  \mean{\xi(x)\xi(x')}_S=\frac{H^3}{4\pi^2}
  \big(1+\mathcal{O}(\sigma^2)\big)\delta(t-t');
\end{equation} 
which is highly divergent for $t=t'$ and has a vanishing characteristic
correlation time, while a smooth window yields a correlation function with
no divergences and a finite correlation time, therefore producing a
colored noise. Namely, choosing
\begin{equation}
\label{window}
  W_\sigma(k\t)=1-e^{-\frac{k^2\t^2}{2\s^2}},
\end{equation}
the two-point correlation function for $r=0$ can be calculated, yielding
\begin{equation}
  \mean{\xi(t)\xi(t')}_S =
  \frac{H^4}{8\pi^2}\frac{1}{\cosh^2(H(t-t'))}+\mathcal{O}(\s^2),
\end{equation}
that asymptotically behaves like $e^{-2H(t-t')}$. Moreover, it is possible 
to show that this asymptotic result is quite general for a wide class
of smooth window functions \cite{colored1}.


\section{Fluctuations}

\label{fluctuations}

The particular solution of the evolution equation \eqref{eqmotoconf} for the 
fluctuations $\dchik$ sourced by the noise field $\xi$ can be expressed in 
terms of the general solutions $\chi_1=\sqrt{k|\t|}J_\nu(k|\t|)$
and $\chi_2=\sqrt{k|\t|}Y_\nu(k|\t|)$ of the homogeneous equation
\eqref{dSeq}. This solution reads
\begin{align}
\label{exact}
  \dchik[\xi](\t) &=\int_{\tin}^\t \!\!d\tilt\, g(k\t,k\tilde\t)
  \,\xik(\tilt),
\end{align}
where
\begin{equation}
  g(k\t,k\tilde\t) = 
  \frac{\chi_1(k\t)\chi_2(k\tilt)-\chi_2(k\t)\chi_1(k\tilt)}
  {\chi_1'(k\tilt)\chi_2(k\tilt)-\chi_2'(k\tilt)\chi_1(k\tilt)}
\end{equation}
and $\tin$ is the beginning of inflation, at which we set the initial 
condition $\dchik(\tin)=0$. 

Keeping this assumption, we now want
to introduce in our system the constraint that at a much later time
$\t_*$ (roughly about 60 e-folds before the end of inflation) we
have no fluctuations in that part of Universe corresponding
to the present observable sky. This is motivated by the fact that
in our treatment all the points we observe today (over which we measure a
substantial homogeneity) were included at $\t_*$ in the same coarse-grained 
domain. It is thus consistent to assume $\t_*$ the complete homogeneity of
the comoving patch of Universe we observe today, all fluctuations on
smaller scales being generated later by the noise term.
In this approach, we conservatively make no assumptions on the behaviour
on larger unobservable scales. 

We are thus led to consider (for a given noise configuration) a 
different solution for the subsequent evolution of the fluctuations, obtained 
as in \eqref{exact} by starting the integration at $\t_*$, when a
new (stochastic) initial condition holds.
In turn, $\dchik[\xi](\t_*)$ is determined again from \eqref{exact} with
the usual vanishing initial condition at $\tin$. 
However, as far as we are dealing with points inside the present observable
Universe, we can skip the stochastic initial conditions $\t_*$ since their
inverse Fourier transform is assumed to vanish. Therefore, in configuration 
space the subsequent evolution of the fluctuations  will only contain
noise modes integrated after $\t_*$.
That is, for relevant $\xx$'s we write
\begin{equation}
  \df(\xx,\t) =
  \int\!\frac{\dk}{(2\pi)^{3/2}} \frac{e^{i\kk\xx}}{a}\!
  \int_{\t_*}^\t \!\!d\tilt\,g(k\t,k\tilde\t)\,\xik(\tilt)
\end{equation}
and
\begin{equation}
  \df(\xx,\t_*) = \!\int\!\frac{\dk}{(2\pi)^{3/2}} \frac{e^{i\kk\xx}}{a_*}\!
  \int_{\tin}^\t \!\!\!\!d\tilt\,g(k\t_*,k\tilde\t)\,\xik(\tilt) \;,
\end{equation}
where the first equation is only valid for scales inside our observed 
patch of the Universe. 

As expected, since the fluctuation $\dfk[\xi]$ is linear in $\xi$, 
at all times we have that
\begin{equation}
  \mean{\df[\xi](\t)}_S=0,
\end{equation}
while the two-point correlation function in $\xx_1$ and $\xx_2=\xx_1+\rr$ 
can be obtained integrating the noise correlation function \eqref{noisecorr}.
We find 
\begin{align}
  C(\rr) &\equiv
  \mean{\df[\xi](\xx_1,\t)\df[\xi](\xx_2,\t)}_S \nonumber \\
  &= \int\!\frac{\dk}{(2\pi)^3}e^{i\kk\rr}
  \frac{\left|I_1(k)\right|^2}{2 k^3},
\end{align}
where
\begin{equation}
\label{int}
  I_1(k)=\frac{\sqrt{2k^3}}{a}\int_{\t_*}^\t \!\!d\tilt\,g(k\t,k\tilde\t)
  (W_\sigma''\chi_\kk+2W_\sigma'\chi'_\kk).
\end{equation}

In order to evaluate this integral, we first change the integration
variable from $\tilt$ to $x=k\tilt$ and apply the relation
\begin{equation}
\label{wronskian}
  J_\nu(x)\frac{d}{dx}Y_\nu(x)- \frac{d}{dx}J_\nu(x)Y_\nu(x)=\frac{2}{\pi x}
\end{equation}
in the denominator of $g(k\t,k\tilde\t)$. Second, we integrate by
parts the $W''_\sigma$ term and and use again \eqref{wronskian}. 
The integral can thus be written as 
\begin{widetext}
\begin{align}
  I_1(k) = H\sqrt{\frac{\pi k^3|\t|^3}{2}}
  \bigg[\left.\frac{\pi}{2}x\, W_\sigma'(x)\Big[Y_\nu(k|\t|)J_\nu(x) 
  - J_\nu(k|\t|)Y_\nu(x)\Big]H_\nu(x)\right|^{k|\t_*|}_{k|\t|}
  + H_\nu(k|\t|) \int^{k|\t_*|}_{k|\t|} \kern -1em dx \, 
  \frac{d}{dx} W_\sigma(x)\bigg],
\label{integr}
\end{align}
\end{widetext}
containing only a boundary term and a trivial integration.

In the same way we can calculate the correlation function evaluated at
$\t_*$, yielding
\begin{align}
  C_*(\rr) &\equiv
  \mean{\df[\xi](\xx_1,\t_*)\df[\xi](\xx_2,\t_*)}_S \nonumber \\
  &= \int\!\frac{\dk}{(2\pi)^3}e^{i\kk\rr}
  \frac{\left|I_2(k)\right|^2}{2 k^3}
\end{align}
where $I_2(k)$ has the same form as $I_1(k)$ but it refers to the time 
interval $[\tin,\t_*]$.
However, the boundary term is now proportional to 
$W_\sigma'(k|\tin|)$ and its contribution to the correlation function 
will be effective only on the scales that crossed the horizon at the 
beginning of inflation. Therefore, since in most realistic models 
$\t_*\gg\tin$, the boundary term in our treatment is completely negligible,
and we have 
\begin{multline}
  I_2(k) = H\sqrt{\frac{\pi k^3|\t_*|^3}{2}} H_\nu(k|\t_*|)   \\
  \Big(W_\sigma(k|\tin|)-W_\sigma(k|\t_*|)\Big).
\end{multline}

Moreover, we can also define the (generally non-vanishing) mixed 
correlation function 
\begin{align}
  M(\rr)  &\equiv
  \mean{\df[\xi](\xx_1,\t)\df[\xi](\xx_2,\t_*)}_S \nonumber \\
  &= \int\!\frac{\dk}{(2\pi)^3}e^{i\kk\rr}
  \frac{\re[I_1(k)I^*_2(k)]}{2 k^3},
\end{align}
involving the scalar field perturbations evaluated at different times.


\section{PROBABILITY DISTRIBUTION}

\label{probabdistr}

We can construct the joint probability distribution 
$P\big[\df_1,\df_2,c]$ that the stochastic variables $\df[\xi](\xx_1)$ 
and $\df[\xi](\xx_2)$ satisfy the constraint $c$ (i.e. they vanish)  
at time $\t_*$ \emph{and} assume the values $\df_1$ and $\df_2$ 
at time $\t$ by taking the statistical average 
\begin{widetext}
\begin{align}
  P\big[\df_1,\df_2,c]
  &= \mean{\prod_{i=1,2}\delta\Big(\df_i-\df[\xi](\xx_i,\t)\Big)
  \delta\Big(\df[\xi](\xx_i,\t_*)\Big)}_S \\
  &= \int\frac{d\al_1d\al_2d\beta_1d\beta_2}{(2\pi)^4}
  e^{-i\sum\al_i \df_i} \int\mathcal{D}[\xi]
  e^{i\sum\al_i\df[\xi](\xx_i,\t)
  +i\sum\beta_i \df[\xi](\xx_i,\t_*)}P[\xi]. \notag
\end{align}
Since the field $\df[\xi]$ is linear in the noise $\xi$, the evaluation of 
the mean consists in a simple Gaussian functional integration yielding 
(repeated indices are summed)
\begin{equation}
\label{joint}
  P\big(\df_1,\df_2,c\big) = 
  \int\frac{d\al_1d\al_2}{(2\pi)^2}
  e^{-i\sum\al_i \df_i-\frac{1}{2}\al_iC(\rr_{ij})\al_j}
  \int\frac{d\beta_1d\beta_2}{(2\pi)^2}
  e^{-\frac{1}{2}\beta_iC_*(\rr_{ij})\beta_j
  -\al_iM(\rr_{ij})\beta_j},
\end{equation}
\end{widetext}
where $\rr_{11}=\rr_{22}=0$ while $\rr_{12}=\rr_{21}=\rr$, and thus the 
result now only involves $2\times2$ symmetric correlation matrices.

According to Bayes theorem, the conditional probability 
$P_c\big(\df_1,\df_2\big)$ is the joint probability 
$P\big(\df_1,\df_2,0,0\big)$ normalized by the probability 
$P_*(0,0)$ of the constraint. The latter can be evaluated exactly
in the same way, taking the mean value over the noise configurations
of the product of two $\delta$-functions constraining the value of the 
fluctuations at $\t_*$. Following the same steps as before, 
the constraint probability reads
\begin{align}
  P_*\big(c\big) 
  &= \int\frac{d\beta_1d\beta_2}{(2\pi)^2}
  e^{-\frac{1}{2}\beta_iC_*(\rr_{ij})\beta_j},
\end{align}
and thus provides the correct normalization needed in order to evaluate 
in \eqref{joint} the Gaussian integration over $\beta_1$ and $\beta_2$. 
With these results, we are now able to compute the conditional two-point 
correlation function
\begin{widetext}
\begin{align}
  \mean{\df(\xx_1)\df(\xx_2)}_c 
  &\equiv\int\df_1\df_2 
  \frac{P\big(\df_1,\df_2,c\big)}{P_*(c)} \notag \\
  &= \int\df_1\df_2\int\frac{d\al_1d\al_2}{(2\pi)^2}
  e^{-i\sum\al_i \df_i-\frac{1}{2}\al_i \left(C(\rr_{ij}) -
  M(\rr_{ik})[C_*(\rr_{kl})]^{-1}M(\rr_{lj})\right)\al_j} \notag \\
  &= C(\rr)-2\frac{M(0)M(\rr)}{C_*(0)+C_*(\rr)}
  + \frac{C_*(0)\left(M(0)-M(\rr)\right)^2}{C_*^2(0)-C_*^2(\rr)}.
\end{align}
\end{widetext}

This result looks rather messy, but it considerably simplifies if we 
take the very reasonable limit $\t_*\gg\tin$ (as it is the case in most
inflationary models). Actually, we can assume that $I_2(k)$ is given by
\eqref{integr} (with $\t_*$ instead of $\t$) neglecting the boundary term,
since this would modify the correlation function only on extremely large
scales (those that crossed the horizon at the beginning of inflation).
We then get
\begin{equation}
  \frac{|I_2(k)|}{H a_*}\sim (k|\t_*|)^{\eps} 
\big(W_\sigma(k|\tin|)-W_\sigma(k|\t_*|)\big),
\end{equation}
which is non-vanishing only for $|\tin|^{-1}\lesssim k\lesssim|\t_*|^{-1}$,
and since the spatial fluctuations are effective for 
$k\gtrsim r^{-1}\gtrsim|\t_*|^{-1}$ and are therefore damped by the cutoff 
introduced by $W_\sigma(k|\t_*|)$ we roughly have
\begin{align}
  C_*(\rr)\sim \frac{H^2}{4\pi^2}
  \int_{|\tin|^{-1}}^{|\t_*|^{-1}}
  \frac{dk}{k} (k|\t_*|)^{2\eps}, 
\end{align}
which is divergent for $|\tin|\rightarrow\infty$ and $\eps\rightarrow 0$. 

However, the dependence on $\tin$ contained in $I_2(k)$ is rapidly saturated
in the mixed correlation function: actually, it enters in $M(\rr)$ only
through the product $I_1(k)I_2^*(k)$, and $I_1(k)$ (which is again obtained
from \eqref{integr} with $\t_*$ instead of $\tin$) significantly differs
from 0 only for $k\gtrsim|\t_*|^{-1}$. Therefore, when integrating over
$\frac{dk}{k}$ the modes responsible for the $\tin$-dependence (\emph{i.e.}
those such that $k\sim|\tin|^{-1}$) are completely damped, $M(\rr)$
is constant in the limit $\tin\ll\t_*$, and in this approximation we get
\begin{equation}
  \mean{\df(\xx_1)\df(\xx_2)}_c\simeq C(\rr).
\end{equation}

The power-spectrum $\spect$ of the fluctuations, defined so that
\begin{equation}
\mean{\df(\xx_1)\df(\xx_2)}_c= \frac{1}{4\pi}\int\dk \;e^{i\kk\rr}
\frac{\spect}{k^3},
\end{equation}
becomes
\begin{equation}
  \spect= \frac{1}{4\pi^2}|I_1(k)|^2.
\end{equation}
For relevant scales and small values of $\sigma$ we can neglect the
boundary term in \eqref{integr}. 
In the small-$\sigma$ limit we therefore recover the standard 
scale-invariant result $\spect=H^2/4\pi^2$.


\section{Curvature Perturbations}

\label{curvature}

We have so far treated the scalar field perturbations generated during 
a de Sitter stage, when the evolution of the scale factor is \emph{a priori} 
fixed independently of the behaviour of the scalar field.
However, if this scalar field is the inflaton, it is the dominating 
component driving the accelerated expansion; 
its perturbations will then in turn modify the energy-momentum tensor, 
inducing fluctuations in the metric and specially 
in the scale factor evolution, via the slow-roll Friedmann equation $
H^2\simeq (8\pi G/3) V(\phi)$. Besides 
scalar field perturbations $\df$, we must now also consider the small 
metric perturbations. 
Both the metric fluctuations and the inflaton perturbation $\df$ will
assume different values depending on the choice of the coordinate frame.
Therefore, in order to avoid this ambiguity and deal only with physical 
degrees of freedom, it is convenient to define the gauge-invariant
comoving curvature perturbation ${\cal R}=\psi+H(\delta\varphi/\dot\phi)$
which measures the intrinsic spatial curvature on hypersurfaces
of constant time \cite{tonireview}. 
Defining in conformal time $z=a^2\phi'/a'$ and
$u=-z\,{\cal R}$, the latter variable  
satisfies the equation of motion 
\begin{equation}
  u''-\nabla^2u-\frac{z''}{z}u=0.
\end{equation}
Expanding the last term to first order in the slow-roll parameters
$\eps= -\dot H/H^2$ and $\eta=V''/3H^2$, one finds
\begin{equation}
  \frac{z''}{z}\simeq\frac{1}{\t^2}\bigg(\nu^2-\frac{1}{4}\bigg),
\end{equation}
where $\nu\simeq\frac{3}{2} + 3\eps - \eta$.

Therefore, in the slow-roll approximation the gauge-invariant normal
modes $u_\kk$ satisfy the same equation of motion \eqref{dSeq}, 
the only difference being in the 
definition of the parameter $\nu$ labelling the solutions.
We can thus apply also to $u$ the stochastic formalism we developed  in Sec. 
\ref{effaction} for the case of a massive scalar field in a pure de Sitter 
background, obtaining
\begin{equation}
  u_\kk'' + \left[k^2 - 
  \frac{1}{\t^2}\left(\nu^2-\frac{1}{4}\right)\right]u_\kk=\xi_\kk.
\end{equation}
We can then apply to $\mathcal{R}$ the results derived for the 
power-spectrum of the perturbations of a test scalar field, 
concluding that also for the curvature perturbation we have
$\mathcal{P}_\mathcal{R}(k)\propto|I_1(k)|^2$. 

More precisely, recalling \eqref{integr} and taking the limit 
$k|\t|\ll \sigma\lesssim 1$  (which is reasonably satisfied on 
cosmological scales) we get
\begin{multline}
  \mathcal{P}_\mathcal{R}(k) = A^2_\mathcal{R} 
   \Big\vert\frac{\pi}{2}k|\t_*|\, 
  J_\nu(k|\t_*|)H_\nu(k|\t_*|)W_\sigma'(k|\t_*|) \\
  + iW_\sigma(k|\t_*|)\Big\vert^2 (k|\t|)^{2\eta-6\eps}.
\end{multline}

If the boundary term in \eqref{integr} can be neglected, 
the power-spectrum becomes
\begin{equation}
  \mathcal{P}_\mathcal{R}(k) =
  A^2_\mathcal{R} W_\sigma^2(k|\t_*|)  (k|\t|)^{2\eta-6\eps}.
\end{equation}
thereby showing, since $W_\sigma^2<1$, the presence of a blue-tilt on the 
largest observable scales. However, in the limit where $\sigma\ll k|\t_*|$
(since $W_\sigma(k|\t_*|)\simeq 1$) we recover the ordinary result
\begin{equation}
  \mathcal{P}_\mathcal{R}(k) =
  A^2_\mathcal{R} (k|\t|)^{2\eta-6\eps}.
\end{equation}

\begin{figure}[t]
  \psfrag{a}{\raisebox{-.25cm}{$k\t_*$}}
  \psfrag{b}[][][1][90]{\kern -14em 
  \raisebox{55pt}{$\mathcal{P}_\mathcal{R}(k)/A^2_\mathcal{R}$}}
  \includegraphics[width=.39\textwidth,height=.31\textwidth]{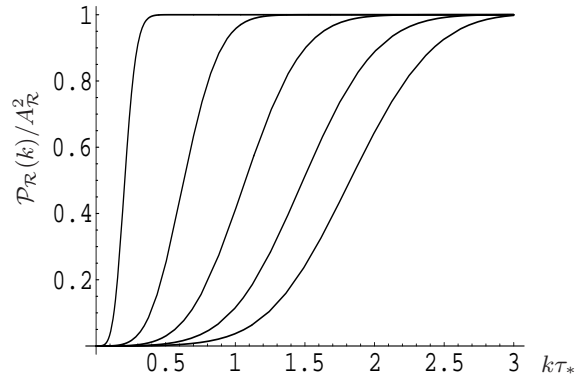}
  \caption{\label{tilt} Conditional power-spectrum obtained with the Gaussian 
  window \eqref{window} in the limit $\t_*/\tin\simeq\infty$, for different 
  values of $\sigma$ varying from 0.1 (left curve) to 0.9 (right curve).}
\end{figure}

For generic values of $\sigma$, we show in Fig.~\ref{tilt} the 
power-spectrum obtained with the Gaussian window function \eqref{window}, 
where the only approximation is that of considering $\tin\ll\t_*$.
For this particular choice of the filter, we observe a blue tilt of the
power-spectrum for $k\lesssim3\sigma a_*H_*$.

This blue tilt stems from the fact that a smooth window function 
does not make a sharp separation in Fourier space but it gradually weighs
the modes, allowing for a small low-frequency contribution to the 
fine-grained part of the field (in term of which the noise is defined)
while depleting modes whose wavelength is immediately smaller than the cutoff
scale.
The colored noise originated from such a window is thus able 
to generate fewer fluctuations than a white noise on scales slightly smaller
than the comoving coarse-graining domain.

\begin{figure*}
\includegraphics[width=.65\textwidth,height=.55\textwidth]{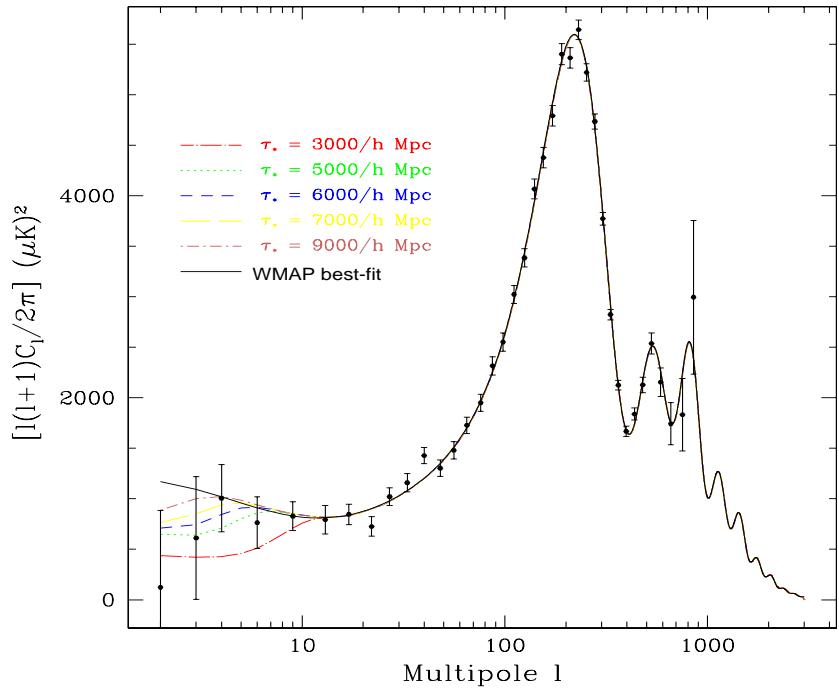}
  \caption{\label{cl}.
Angular power spectra of the CMB temperature anisotropies obtained with
the Gaussian window \eqref{window} and different values of the constraint time
$\t_*$. The cosmological parameters used in the computation of the
radiation transfer functions are $\Omega_b = 0.046$, $\Omega_{CDM} = 0.224$,
$\Omega_\Lambda = 0.730$ and $h = 0.72$, corresponding to the best fit of the 
WMAP data \cite{hinshaw}. Dots represent the WMAP Binned Combined TT Power
Spectrum, obtained from the LAMBDA website: http://lambda.gsfc.nasa.gov/.}
\end{figure*}

As a consequence, since we impose the constraint that in our comoving 
patch of Universe the fluctuations can grow only after $\t_*$, the scales 
that are leaving the horizon in the following few Hubble times receive less
``random kicks'' before freezing out than in the white-noise case. 
Therefore, the power-spectrum is a function of $k$ smoothly interpolating  
between the values 0 and 1 it assumes for small and large $k$, respectively.

This power-spectrum can be used to calculate the CMB multipoles predicted
by a specific choice of the window function $W$. Quite generally, we will
expect to find a suppression of the lowest multipole, which are sensitive 
to a modification of the power-spectrum on this very large scales.
However, in order to quantify this suppression one needs to choose
the shape of the window function and the precise time $\t_*$ at which the
constraint is set.
For the Gaussian window \eqref{window}, the predicted CMB multipoles
are plotted in Fig.~\ref{cl} for different values of 
$\t_*$. The cosmological parameter used to compute (with CMBfast 
\cite{CMBfast}) the radiation transfer functions are the best fit of the 
\emph{WMAP} data \cite{WMAP}.
The significance of the low multipoles suppression varies depending on the
choice of the constraint time: the effect is strong for
$\t_*\simeq3000h^{-1}$ Mpc (corresponding to the present horizon size),
while it becomes fainter for earlier values of $\t_*$ and it is practically
absent for $\t_*\simeq9000h^{-1}$ Mpc. 

\section{Conclusions}
\label{conclu}

In this paper we calculated the power-spectrum
of the comoving curvature perturbation ${\cal R}$ in the framework
of stochastic inflation with colored noise.
We can conclude that a careful analysis of the stochastic behaviour
of the fluctuations, for a physically plausible choice of the
window function $W_\sigma$ (therefore excluding the step function), leads 
to a curvature power-spectrum showing a relevant blue tilt for scales such 
that $k\sim\sigma a_*H_*$, corresponding to physical lengths about 
$\sigma^{-1}$ times greater than the present Hubble radius. 

The possibility of finding a blue tilt on observable scales will then
depend on the size of the coarse-graining domain at which 
thehomogeneity constraint is set. If $\sigma a_*H_*$ is of the order 
of $a_0 H_0$ (today's Hubble radius), then also the biggest scales in our 
observable Universe get tilted, while if it is much smaller this effect might 
be completely unobservable. 
We computed the predicted angular power-spectrum $C_l$'s of CMB temperature
anisotropies for a particular choice of the colored-noise window function, 
finding a suppression of the lowest multipoles whose importance depends on the 
precise value of $\t_*$: this effect actally becomes more and more 
evident as $\t_*$ approaches $3000h^{-1}$ Mpc (the size of the present
Hubble radius), whike it is practically absent for $\t_*\simeq9000h^{-1}$
Mpc. 
This is an interesting feature because the recent measurements of the
CMB anisotropy performed by \emph{WMAP} seem to give evidence of a lack 
of power on the largest observed scales with respect to the predictions
of the standard inflationary scenario \cite{hinshaw}.
In our approach, this deviation from the standard result of an almost 
scale-invariant perturbation spectrum is achieved without any new physical 
input in the theory, being merely a consequence of the choice of initial 
conditions once some over-simplifying approximations (namely, the
white-noise choice) are removed. We find it encouraging that
no introduction of new physical ingredients is needed to explain
the anomalous behaviour of the lower multipoles of CMB 
anisotropies.

\section*{ACKNOWLEDGMENTS}

MM wishes to thank the University of Milano-Bicocca and the Institut 
d'Astrophysique de Paris for hospitality, and J\'er\^ome Martin for
interesting and useful discussions. We acknowledge the use of the 
Legacy Archive for Microwave Background Data Analysis (LAMBDA). 
Support for LAMBDA is provided by the NASA Office of Space Science. 

\vspace{\stretch{1}}


\newpage


\begin{thebibliography}{50}

\bibitem{lrreview} D.~H.~Lyth and A.~Riotto,
Phys.\ Rept.\  {\bf 314}, 1 (1999)
[arXiv:hep-ph/9807278].
\bibitem{tonireview} A.~Riotto,
Lectures given at ICTP Summer School on Astroparticle Physics and Cosmology, 
Trieste, Italy, 17 Jun - 5 Jul 2002, published in Trieste 2002, 
{\it Astroparticle physics and cosmology},  317-413; 
arXiv:hep-ph/0210162.
\bibitem{early} A.~A.~Starobinsky, in  {\it Field Theory, Quantum 
Gravity and Strings}, H.~J~ De Vega, N.~Sanchez Eds., 107 (1986);
A.~S.~Goncharov, A.~D.~Linde and V.~F.~Mukhanov, 
Int.\ J.\ Mod.\ Phys.\ A {\bf 2}, 561 (1987);
S.~J.~Rey, Nucl.\ Phys.\ B {\bf 284}, 706 (1987);
K.~I.~Nakao, Y.~Nambu and M.~Sasaki, 
Prog.\ Theor.\ Phys. {\bf 80}, 1041 (1988);
Y.~Nambu and M.~Sasaki, 
Phys.\ Lett.\ B {\bf 219}, 240 (1989);
Y.~Nambu, 
Prog. Theor. Phys. {\bf 81}, 1037 (1989);
A.~D.~Linde, D.~A.~Linde and A.~Mezhlumian,
Phys.\ Rev.\ D {\bf 49}, 1783 (1994).

\bibitem{decoherence}
D.~Polarski and A.~A.~Starobinsky,
Class.\ Quant.\ Grav.\  {\bf 13}, 377 (1996)
S.~Habib, 
Phys.\ Rev.\ D {\bf 46}, 2408 (1992).
E.~Calzetta and B.~L.~Hu,
Phys.\ Rev.\ D {\bf 52}, 6770 (1995).
M.~Bellini, H.~Casini, R.~Montemayor and P.~Sisterna, 
Phys.\ Rev.\ D {\bf 54}, 7172 (1996).
A.~Matacz, 
Phys.\ Rev.\ D {\bf 55}, 1860 (1997).
C.~Kiefer, J.~Lesgourgues, D.~Polarski and A.~A.~Starobinsky,
Class.\ Quant.\ Grav.\  {\bf 15}, L67 (1998).

\bibitem{fokkerplanck} H.~Risken, \emph{The Fokker-Planck Equation: 
methods of solution and applications}, Springer-Verlag, Berlin (1984);
S.~Chandrasekhar, Rev.\ Mod.\ Phys. {\bf 15}, 1 (1943).

\bibitem{constraint} 
D.~S.~Salopek and J.~R.~Bond,
Phys.\ Rev.\ D {\bf 43}, 1005 (1991);
I.~Yi, E.~T.~Vishniac and S.~Mineshige,
Phys.\ Rev.\ D {\bf 43}, 362 (1991);
I.~Yi and E.~T.~Vishniac,
Phys.\ Rev.\ D {\bf 45}, 3441 (1992);
I.~Yi and E.~T.~Vishniac,
Astrophys.\ J.\ Suppl. {\bf 86}, 333 (1993);
H.~M.~Hodges, 
Phys. Rev. Lett.  {\bf 64}, 1080 (1990);
S. Mollerach, S. Matarrese, A. Ortolan and F. Lucchin, 
Phys. Rev. D {\bf 44}, 1670 (1991);
L.~Kofman, G.~R.~Blumenthal, H.~Hodges and J.~R.~Primack,
ASP Conf.\ Ser.\  {\bf 15}, 339 (1991).
S.~Matarrese, in {\it 6th Moriond Astrophysics Mtg.: 
The Early Universe and Cosmic Structures}, J.~M.~Alimi {\it et al.} 
Eds., 21 (1990).

\bibitem{stochNG}
A.~Gangui, F.~Lucchin, S.~Matarrese and S.~Mollerach,
Astrophys.\ J.\  {\bf 430}, 447 (1994); 
L.~M.~Wang and M.~Kamionkowski,
Phys.\ Rev.\ D {\bf 61}, 063504 (2000);
G.~I.~Rigopoulos and E.~P.~S.~Shellard,
arXiv:astro-ph/0405185.


\bibitem{colored1}
S.~Winitzki and A.~Vilenkin, 
Phys.\ Rev.\ D {\bf 61}, 084008 (2000)



\bibitem{colored2} 
B.~L.~Hu, J.~P.~Paz and Y.~H.~Zhang,
Phys.\ Rev.\ D {\bf 45}, 2843 (1992).
H.~Casini, R.~Montemayor and P.~Sisterna, Phys.\ Rev.\ 
D{\bf 59}, 063512 (1999);
B.~L.~Hu, J.~P.~Paz and Y.~H.~Zhang,
in \emph{Chateau du Pont d'Oye 1992, Proceedings, The origin of structure in 
the universe}, E. Gunzig, P. Nardone Eds., 227-251 (1992) 
[arXiv:gr-qc/9512049].



\bibitem{MMR}
S.~Matarrese, M.~A.~Musso and A.~Riotto,
JCAP 05 (2004) 008.



\bibitem{WMAP} 
D.~N.~Spergel {\it et al.},
Astrophys.\ J.\ Suppl.\  {\bf 148}, 175 (2003).



\bibitem{COBE}
C.~L.~Bennett {\it et al.},
Astrophys.\ J.\  {\bf 464}, L1 (1996).



\bibitem{lowmultstat}
M.~Tegmark, A.~de Oliveira-Costa and A.~Hamilton,
Phys.\ Rev.\ D {\bf 68}, 123523 (2003);
A.~de Oliveira-Costa, M.~Tegmark, M.~Zaldarriaga and A.~Hamilton,
Phys.\ Rev.\ D {\bf 69}, 063516 (2004);
G.~Efstathiou,
Mon.\ Not.\ Roy.\ Astron.\ Soc.\  {\bf 346}, L26 (2003);
G.~Efstathiou,
Mon.\ Not.\ Roy.\ Astron.\ Soc.\  {\bf 348}, 885 (2004);
P.~Bielewicz, K.~M.~G\'orski and A.~J.~Banday,
arXiv:astro-ph/0405007.



\bibitem{blue-tilt1}
L.~R.~Abramo and L.~J.~Sodr\'e,
arXiv:astro-ph/0312124.



\bibitem{blue-tilt2}
Y.~P.~Jing and L.~Z.~Fang,
Phys.\ Rev.\ Lett.\  {\bf 73}, 1882 (1994)
C.~R.~Contaldi, M.~Peloso, L.~Kofman and A.~Linde,
JCAP {\bf 0307}, 002 (2003);
S.~L.~Bridle, A.~M.~Lewis, J.~Weller and G.~Efstathiou,
Mon.\ Not.\ Roy.\ Astron.\ Soc.\  {\bf 342}, L72 (2003)
G.~Efstathiou,
Mon.\ Not.\ Roy.\ Astron.\ Soc.\  {\bf 343}, L95 (2003);
B.~Feng and X.~Zhang,
Phys.\ Lett.\ B {\bf 570}, 145 (2003)
A.~Melchiorri and L.~Mersini-Houghton,
arXiv:hep-ph/0403222;
T.~Moroi and T.~Takahashi,
Phys.\ Rev.\ Lett.\  {\bf 92}, 091301 (2004);
Y.~S.~Piao, B.~Feng and X.~m.~Zhang,
Phys.\ Rev.\ D {\bf 63}, 084520 (2000);
Y.~S.~Piao, S.~Tsujikawa and X.~M.~Zhang,
arXiv:hep-th/0312139.



\bibitem{noneqFT} 
J. Schwinger, J. Math. Phys. {\bf 2}, 407 (1961);
R. D. Jordan, Phys. Rev. D {\bf 33}, 444 (1987).



\bibitem{influence} 
M.~Morikawa, Phys.\ Rev.\ D {\bf 42}, 1027 (1990);
B.~L.~Hu and S.~Sinha,
Phys.\ Rev.\ D {\bf 51}, 1587 (1995).

\bibitem{CMBfast}
U.~Seljak and M.~Zaldarriaga,
Astrophys.\ J.\  {\bf 469}, 437 (1996)

\bibitem{hinshaw}
G.~Hinshaw {\it et al.},
Astrophys.\ J.\ Suppl.\  {\bf 148}, 135 (2003)



\end{thebibliography}
\end{document}